\documentclass[prl,showpacs,preprintnumbers,amsmath,amssymb,twocolumn]{revtex4}

\usepackage{graphicx}
\usepackage{dcolumn}
\usepackage{bm}

\begin{document}

\title{Defects in  CrAs and related compounds: a route to half-metallic ferrimagnetism}

\author{I. Galanakis$^1$}\email{galanakis@upatras.gr} \author{K. \"Ozdo\~gan$^2$}\email{kozdogan@gyte.edu.tr}
\author{E. \c Sa\c s\i
o\~glu$^{3,4}$}\email{e.sasioglu@fz-juelich.de}
\author{B. Akta\c s$^2$}

\affiliation{$^1$ Department of Materials Science, School
of Natural Sciences, University of Patras,  GR-26504 Patra, Greece\\
$^2$ Department of Physics, Gebze Institute of Technology, Gebze,
41400, Kocaeli, Turkey\\
$^3$ Institut f\"ur Festk\"orperforschung, Forschungszentrum
J\"ulich, D-52425 J\"ulich, Germany\\
$^4$ Fatih University,  Physics Department, 34500, B\" uy\" uk\c
cekmece,  \.{I}stanbul, Turkey}

\date{\today}

\begin{abstract}
Half-metallic ferrimagnetism is crucial for spintronic
applications with respect to ferromagnets due to the lower stray
fields created by these materials. Studying the effect of defects
in CrAs and related transition-metal chalcogenides and pnictides
crystallizing in the zinc-blende structure, we reveal that the
excess of the transition-metal atoms leads to half-metallic
ferrimagnetism. The surplus of these atoms are
antiferromagnetically coupled to the transition-metal atoms
sitting at the perfect lattice sites. The needed condition to
achieve half-metallic ferrimagnetism is to prevent the migration
of the $sp$ atoms to other sites and the atomic swaps.
\end{abstract}


\maketitle

The scientific research on the spintronic materials has exploded
during the last years \cite{Zutic}. Half-metallic ferromagnets
like Heusler alloys (NiMnSb \cite{deGroot,GalanakisHalf}) or some
oxides (CrO$_2$, Fe$_3$O$_4$, LSMO \cite{Soulen}) play a central
role since they can be used to maximize the efficiency of such
devices due the almost perfect spin-polarization at the Fermi
level. The ideal case for applications would be a half-metallic
antiferromagnet (also known as fully-compensated ferrimagnet) like
CrMnSb, where the majority spin and minority spin densities of
states are not identical, as for common antiferromagnets. Such a
compound would be a perfectly stable spin-polarized electrode in a
junction device. And moreover if used as a tip in a spin-polarized
STM, it would not give rise to stray flux, and hence would not
distort the domain structure of the soft-magnetic systems to be
studied. Unfortunately, CrMnSb does not crystallize in the
structure which is theoretically predicted to present
half-metallicity. In the absence of half-metallic
antiferromagnets, the study of ferrimagnets becomes
 important. Van Leuken and de Groot have shown that doping of the
semiconductor FeVSb results in a half-metallic ferrimagnet
\cite{Leuken}. Also some other Heusler compounds like FeMnSb
\cite{GalanakisHalf} and Mn$_2$VAl \cite{Kemal,GalanakisFull}  are
predicted to be half-metallic ferrimagnets but no concrete
experimental results exist.

Since the discovery of half-metallic ferromagnetism in thin films
of CrAs in the zinc-blende structure by the group of Akinaga in
2000 \cite{Akinaga2000}, the transition-metal chalcogenides and
pnictides have attracted considerable attention. Experimentally
several such compounds have been grown in thin-films, multilayers
or dots structures \cite{experiments}. A large number of ab-initio
calculations have also contributed to the understanding of the
basic physics of these alloys \cite{MavropoulosZB,calculations}.
The gap in the minority-spin band arises from the hybridization
between the p-states of the $sp$ atom and the triple-degenerated
$t_{2g}$ states  of the transition-metal \cite{MavropoulosZB}. As
a result the total spin-moment, $M_t$, follows the Slater-Pauling
(SP) behavior being equal in $\mu_B$ to $Z_t-8$ where $Z_t$ the
total number of valence electrons in the unit cell
\cite{MavropoulosZB}. Some of the most recent results include the
study of the exchange bias in ferro-/antiferromagnetic interfaces
\cite{Nakamura2006}, the study of the stability of the zinc-blende
structure \cite{Xie2003}, and the study of dynamical correlations
\cite{Chioncel2006}.

States at the interfaces of these compounds with semiconductors
seem not to affect the half-metallicity \cite{Interfaces}. On the
other hand temperature effects play a more crucial role
\cite{MavropoulosTemp}. At low temperatures the interaction of the
electrons with magnons leads to non-quasiparticle excitations in
the minority gap above the Fermi level \cite{Chioncel2005}, while
at higher temperatures spin wave excitations lead to new states in
the gap above and below the Fermi level \cite{Skomski2002}. Except
interface states and temperature, the third main effect which can
destroy half-metallicity is the appearance of defects and
disorder. Our initial aim in this manuscript is to study their
consequences on the electronic and magnetic properties of CrAs and
related compounds in the zinc-blende (zb) structure. To have a
global view of the behavior of defects, we have included six
different compounds in our study: CrAs and its isovalence CrSb,
CrSe and CrTe with one valence electron more than CrAs, and
finally VAs and MnAs. The electronic structure calculations are
performed using the full--potential nonorthogonal local--orbital
minimum--basis band structure scheme (FPLO) and disorder has been
simulated using the coherent potential approximation
\cite{koepernik}. We have used the theoretical equilibrium lattice
constants for which the perfect compounds are half-metallic
ferromagnets \cite{lattices}. Investigating the properties of
antisites in these alloys where there is an excess of either the
transition metal or the $sp$ atom, \textit{e.g.}
Cr$_{1+x}$As$_{1-x}$, we show that in most cases the half-metallic
ferromagnetism is destroyed, but the excess of Cr leads to the
even more interesting half-metallic ferrimagnets. A study of
atomic swaps as well as Cr antisites at the vacant sites reveals
that they destroy completely the half-metallicity.

\begin{figure}
\includegraphics[scale=0.58]{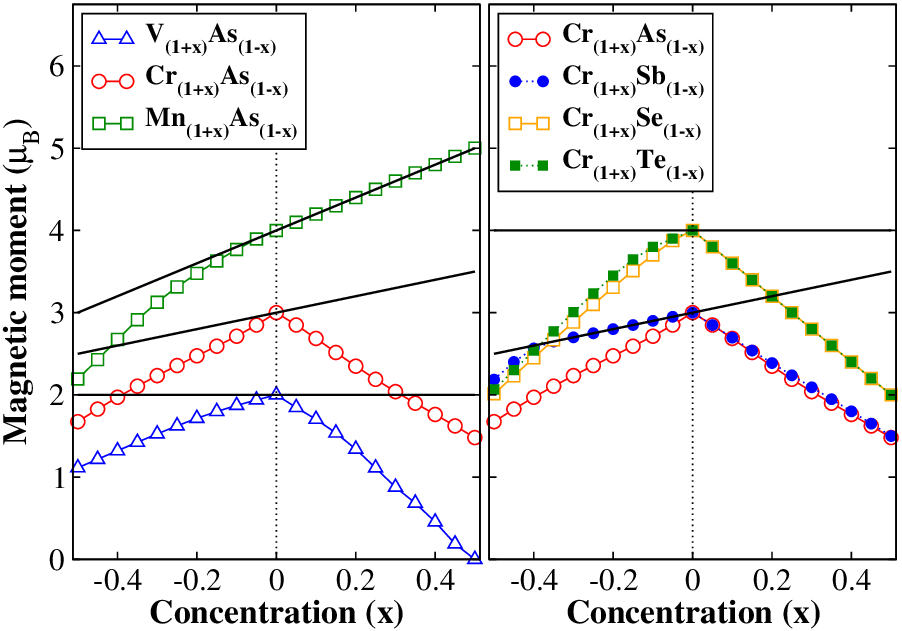}
\caption{(Color online) Total spin magnetic moment as a function
of the concentration $x$ for the studied T$_{1+x}$Z$_{1-x}$
compounds. Note that positive values of $x$ correspond to excess
of the transition metal atom (T) while negative values of $x$ to
excess of the $sp$ atom (Z). The solid black lines represent the
Slater-Pauling behavior for the ideal compounds which they cross
at $x=0$. \label{fig1}}
\end{figure}

Prior to the presentation of our results we have to shortly
describe the zinc-blende structure (zb). The unit cell is that of
a f.c.c. lattice with four atoms per unit cell. The A site located
at $(0\: 0\: 0)$ in Wyckoff coordinates is occupied by the Cr
atoms in the case of CrAs and the B site $(0.25\: 0.25\: 0.25)$ by
As atoms. The C site at $(0.5\: 0.5\: 0.5)$ and D site at $(0.75\:
0.75\: 0.75)$ are vacant and have the same symmetry as the A and B
sites, respectively, rotated by 90$^o$. The need to introduce the
vacant sites in the description of the zb structure arises from
possible migration of Cr atoms towards these sites.

The first case of possible defects under study is when either Cr
or As atoms migrate at antisite positions occupied by the other
chemical element without disturbing the vacant sites. In Fig.
\ref{fig1} we have gathered the total spin moment in the unit cell
as a function of the concentration for all six studied compounds.
In the case of Cr$_{1+x}$As$_{1-x}$ positive values of the
concentration $x$ correspond to Cr-excess and thus the creation of
Cr antisites while negative values of $x$ to the creation of As
antisites. We have used a step of 0.05 and scanned all the region
from $x=-0.5$ up to $x=0.5$. Comparing our results with the ideal
SP behavior of the ideal half-metallic ferromagnets represented by
the solid lines we find that except the case of Mn-antisites in
MnAs and Sb antisites in CrSb, the total spin moment is far away
from the ideal values; a clear sign that half-metallic
ferromagnetism is lost.

\begin{table}
\caption{Total and atom-resolved spin magnetic moments for the
compounds under study. As "imp" we denote the atoms which are
located at antisite positions. Note that the atomic moments have
been scaled to one atom.} \label{table1}
\begin{ruledtabular}
 \begin{tabular}{lccccc}
Compound & $x$ & Total  &  T (Cr,V,Mn)  & Z (As,Sb,Se,Te) & T imp \\
Cr$_{1+x}$As$_{1-x}$  & 0    &  3.0  & 3.4 &  -0.4 & -- \\
                      & 0.1  &  2.7  & 3.3 &  -0.4 & -2.8 \\
                      & 0.2  &  2.4  & 3.2 &  -0.4 & -2.9 \\
Cr$_{1+x}$Sb$_{1-x}$  & 0    &  3.0  & 3.5 &  -0.5 & -- \\
                      & 0.1  &  2.7  & 3.5 &  -0.5 &  -3.3 \\
                      & 0.2  &  2.4  & 3.5 &  -0.5 &  -3.4 \\
Cr$_{1+x}$Se$_{1-x}$  & 0    &  4.0  & 4.2 &  -0.2 & -- \\
 & 0.1   &  3.6   &  4.1 &  -0.2 & -3.1\\
 & 0.2   &  3.2  &  4.0 &  -0.2 &   -3.1 \\
Cr$_{1+x}$Te$_{1-x}$ &  0 & 4.0 & 4.2 & -0.2 & -- \\
 & 0.1  &   3.6 &     4.1 & -0.2 &    -3.3\\
 & 0.2  &   3.2 &     4.0 & -0.2 &    -3.3  \\
V$_{1+x}$As$_{1-x}$ &  0 &      2.0 &      2.3 & -0.3 & -- \\
&  0.1 &    1.7 &     2.0 &     -0.2 & -1.1 \\
 &   0.2 &    1.3 &     1.7 &     -0.2 & -1.1 \\
Mn$_{1+x}$As$_{1-x}$  & 0  & 4.0 &    4.3&  -0.3 & -- \\
 & 0.1 &     4.2    &  4.2 &      -0.3 &   3.3\\
 & 0.2 &     4.4    &  4.0  &    -0.4   &   3.3
\end{tabular}
\end{ruledtabular}
\end{table}

\begin{figure}
\includegraphics[scale=0.58]{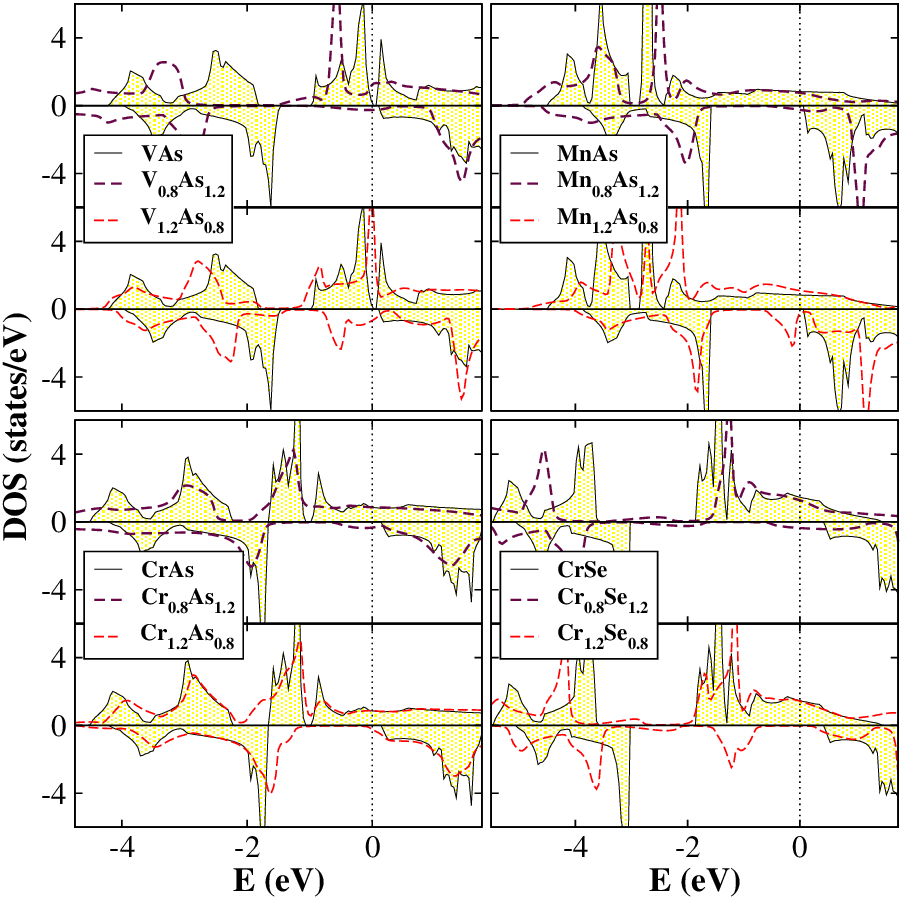}
\caption{(Color online) Density of states (DOS) for the studied
compounds. Positive DOS values correspond to the majority-spin
electrons and negative to the minority ones. The Fermi level is
scaled to the zero of the energy axis. CrSb and CrTe compounds
present DOS similar to CrAs and CrSe, respectively, with the Fermi
level near the middle of the gap.} \label{fig2}
\end{figure}

To understand this behavior we have drawn in Fig. \ref{fig2} the
total DOS for some selected cases and in  Table \ref{table1} we
have gathered the spin magnetic moments for two different values
of the concentration $x$.  When there is an excess of the $sp$
atom  (negative values of $x$) the effect on the electronic
structure is milder. Some of the $p$ states which were located at
the non-occupied minority-spin antibonding states move lower in
energy due to the change in the Coulomb repulsion and are now
crossing the Fermi level. These states are present in all cases.
For CrSb where the Fermi level is deep in the gap, the compound
keeps its half-metallicity up to about $x=-0.4$. The impurity $sp$
atoms at the antisite positions have very small spin moments
parallel to the ones of the transition-metal atoms.

The case of transition-metal antisites at the B site occupied by
the $sp$ atoms is more interesting. In CrAs each Cr atom sitting
in the A-sublattice has four As atoms as first neighbors. When we
create an excess of Cr atoms, these impurities take the place of
As atoms occupying the B-sublattice of the zb structure. Cr(Mn)
atoms are well known to demonstrate ferromagnetic or
antiferromagnetic coupling depending on the distance between
neighboring Cr(Mn) atoms. The distance for the transition from
antiferromagnetic to ferromagnetic coupling is smaller for Mn than
Cr. A similar effect but less intense is also present for the V
atoms although V has a less pronounced magnetic behavior than Cr
and Mn.

When we create an excess of Mn in MnAs, the Mn impurities sitting
at the B-site are ferromagnetically coupled between them since the
intra-sublattice interactions are positive as was discussed in the
paper by Sa\c s\i o\~glu \textit{et al.} on the exchange
interactions in these compounds \cite{Sasioglu-Gala}. The two
sublattices are ferromagnetically coupled and as shown in Table
\ref{table1} all Mn atoms have positive spin-moments. Mn impurity
atoms at B sites have a tetrahedral symmetry with four Mn atoms at
the  A sites as first neighbors. Thus the $d-d$ hybridization for
Mn-impurities is larger and the minority double-degenerated $e_g$
states are occupied (new minority peak just below the Fermi level
in Fig. \ref{fig2}) while all five majority $d$-states are
occupied showing a spin moment of around 3 $\mu_B$. On the other
hand Mn at the perfect sites have now both As and Mn atoms as
first neighbors, the hybridization effects are larger and their
spin moment is smaller with respect to the perfect compound. Thus
the exchange interactions in MnAs are such that as shown in Fig.
\ref{fig1} the half-metallic ferromagnetism is conserved for the
case of Mn antisites at the sublattice occupied by As.

\begin{figure}
\includegraphics[scale=0.58]{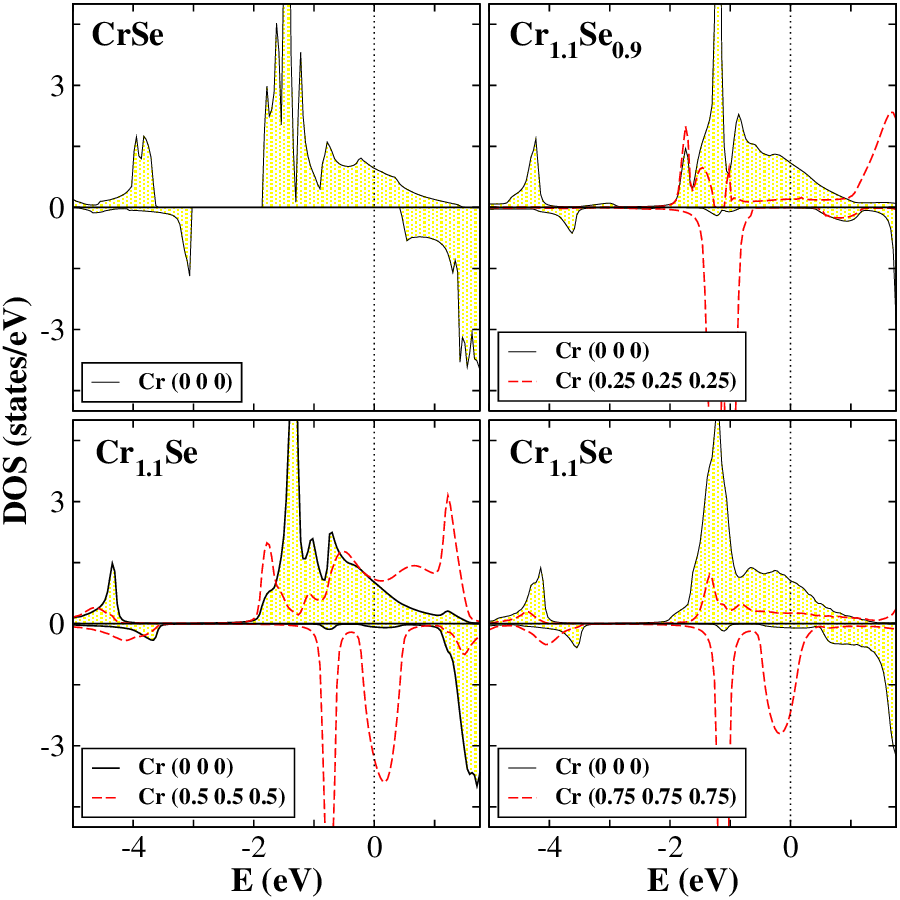}
\caption{(Color online) Cr-resolved DOS for the ideal CrSe
compound (upper left panel), for Cr at the ideal A site  and Cr
impurity at a B site occupied by Se in the perfect compound in the
case of Cr$_{1.1}$Se$_{0.9}$ (upper right panel). In the lower
panels the case of Cr antisites at vacant C (left) and D (right)
sites. \label{fig3}}
\end{figure}

The Cr impurity atoms at the B sites, contrary to the Mn atoms,
are antiferromagnetically coupled to the Cr atoms at the A sites
due to the small distance between the sites at the A and B
sublattices. This phenomenon destroys the ferromagnetism but not
the half-metallicity.  The $d$-states of the Cr impurities are
well localized and the minority-spin (now minority is defined by
Cr atoms at the A site) gap still exists. This is illustrated in
Fig. \ref{fig3} where we present the Cr-resolved DOS. In the upper
left panel is the perfect CrSe case and in the upper right panel
the case of Cr antisites at the B sublattice (Cr$_{1.1}$Se$_{0.9}$
alloy). The Cr impurity atoms in Cr$_{1.1}$Se$_{0.9}$ have all
five bonding minority states occupied as well as  the majority
$e_g$ states and thus spin magnetic moments of around $-3\:
\mu_B$. The minority bonding and antibonding states are separated
by a gap keeping the half-metallic character of the compound.  As
we increase the concentration in Cr impurities at the B sites the
unoccupied minority states of the Cr impurities start forming a
broader band and they approach the Fermi level. No general
statement can be drawn from the total spin magnetic moments since
half-metallic ferrimagnets do not follow the SP behavior and we
have to look the total DOS for several dense values of the
concentration $x$ to identify where the half-metallicity is lost
for these compounds; \textit{e.g.} as shown in Fig. \ref{fig2} for
CrAs  for $x$=0.2 the half-metallicity is lost while for CrSe the
case for $x$=0.2 is half-metallic. Our investigation showed that
the Cr compounds are half-metallic up to concentrations $x$=0.15
for CrAs, 0.2 for CrSb, 0.5 for CrSe and 0.5 for CrTe. We should
also note that the width of the gap is now much smaller since
impurity states occur in the low-energy part of the gap.

For the VAs compound, shown in Fig. \ref{fig2},  the V impurity
atoms are antiferromagnetically coupled to the V atoms at the
ideal sites and the physics of these systems is similar to the
Cr-based compounds. But even for $x$=0.05 the impurity states
cross the Fermi level and although a gap exists the
half-metallicity is lost.

\begin{figure}
\includegraphics[scale=0.55]{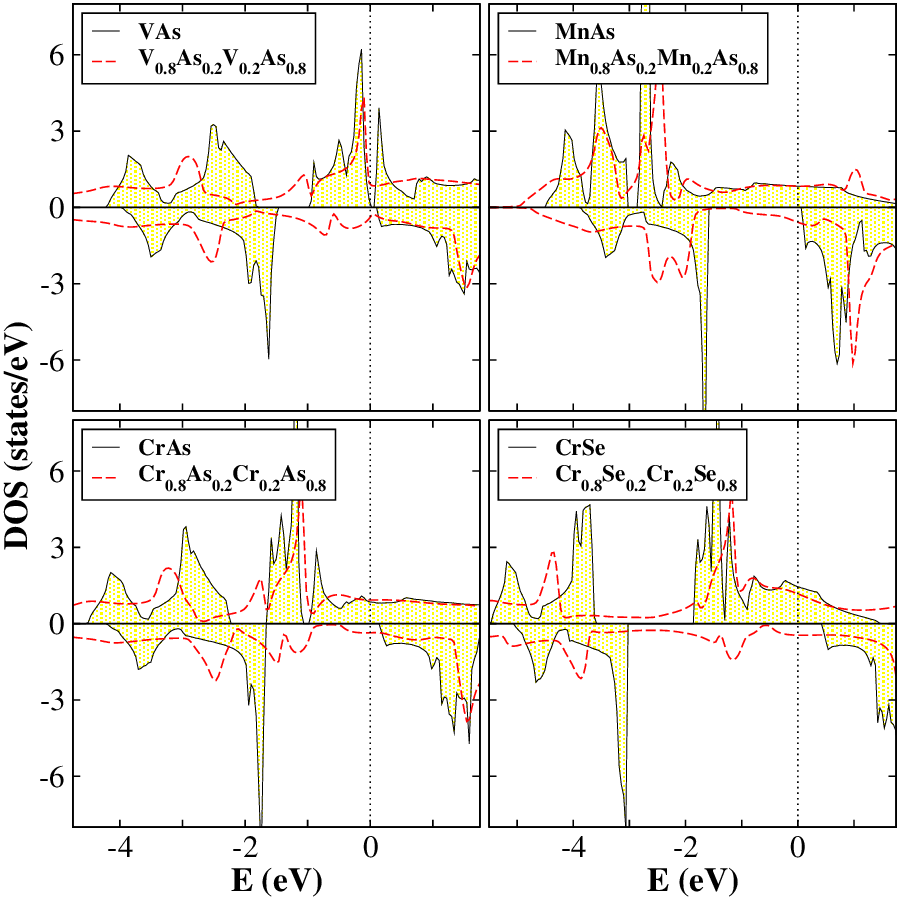}
\caption{(Color online) DOS for the case of the atomic swaps where
20 \%\ ($x=0.2$) of the transition-metal and $sp$ atoms have
exchanged sites.  \label{fig4}}
\end{figure}

We have also studied the case of the creation of Cr antisites at
the vacant sites (lower panel is Fig. \ref{fig3}) and the atomic
swaps (Fig. \ref{fig4}). Cr antisites at the vacant sites
completely destroy half-metallicity since the Fermi level now
crosses the minority triple-degenerated $t_{2g}$ states. But these
antisites are expected to have high formation energies since Fermi
level is now located at a peak of the minority DOS. Atomic swaps
mean that also $sp$ atoms move to the A-sites and the effect is
even more intense than in the case of simple $sp$-antisites
presented above. Even for small values corresponding to the dilute
limit the impurity states destroy completely the minority-spin
gap. The intensity of these states increases rapidly with the
percentage of swaps and the total spin-moment decreases faster
than in the case of simple $sp$ antisites.

We have studied the effect of antisites and atomic swaps in the
case of the transition-metal pnictides and chalcogenides
crystallizing in the zb structure. The $sp$ antisites and the
atomic swaps destroy the half-metallicity. Mn antisites in MnAs
keep the half-metallic ferromagnetic character of the perfect
compounds. The Cr-impurities in the case of antisites at the
sublattice occupied by the $sp$ atoms couple antiferromagnetically
to the existing Cr atoms at the ideal sites and destroy
ferromagnetism. But these compounds stay half-metallic for large
concentration of antisites. Cr antisites at the vacant sites on
the other hand completely destroy half-metallicity. Thus we have
presented an alternative way to create half-metallic ferrimagnets
for realistic spintronic application by simply introducing
Cr-antisites in existing experimental structures based on CrAs or
CrSb. We expect these results to stimulate further interest in
both the theoretical and experimental research in the emerging
field of magnetoelectronics.


\end{document}